\documentclass[prl,aps,twocolumn,showpacs,superscriptaddress]{revtex4}
\usepackage{amssymb,amsmath,array,graphicx,epsfig}

\begin{document}

\title{Theory of the asymmetric ripple phase in achiral lipid membranes}

\author{Md. Arif Kamal}
\author{Antara Pal}
\author{V. A. Raghunathan}
\affiliation{Raman Research Institute, C V Raman Avenue, Bangalore 560 080, India}
\author{Madan Rao}
\affiliation{Raman Research Institute, C V Raman Avenue, Bangalore 560 080, India}
\affiliation{National Centre for Biological Sciences (TIFR), GKVK Campus, Bangalore 560 065, India}

%\date{\today}

\begin{abstract}

We present a phenomenological  theory of phase transitions in achiral lipid membranes 
in  terms of two coupled order parameters -- a scalar order parameter describing {\it lipid chain melting}, 
and a vector order parameter describing the {\it tilt of the hydrocarbon chains} below the
chain-melting transition.  Existing theoretical models fail to account for all the observed features of the 
phase diagram, in particular the detailed microstructure of the {\it asymmetric ripple phase}
lying between the fluid and the tilted gel phase.
In contrast, our two-component theory reproduces all the salient structural features of the ripple phase, 
providing a unified description of the phase diagram and microstructure.

\end{abstract}

\maketitle

Phospholipids self-assemble in water to form a rich
variety of spatially modulated phases~\cite{Tardieu73}.  The simplest of these is the 1-dimensionally modulated fluid
lamellar phase ($L_\alpha$) consisting of periodic stacks of lipid bilayer membranes separated by water, where the
hydrocarbon chains are floppy with liquid-like in-plane order.
Changing the temperature or water
content induces a sequence of symmetry breaking transitions characterized by unique microstructures.

On reducing the temperature below the chain melting (main) transition ($T_m$), the $L_\alpha$ phase of
phosphatidylcholines (PCs)  transforms to a gel phase ($L_{\beta^{\prime}}$), characterized by fully-stretched
$all-trans$ chains which are tilted with respect to the bilayer normal ~\cite{Smith90,Sun94,MacIntosh80}. In addition, 
an {\it asymmetric ripple phase} ($P_{\beta^{'}}$) is found to occur in between the $L_\alpha$ and
$L_{\beta^{'}}$ phases in many PCs at high water content ~\cite{Tardieu73,Smith90,Janiak79}.

\begin{figure}[b]
\begin{center}
\includegraphics[width= 65 mm,angle=270] {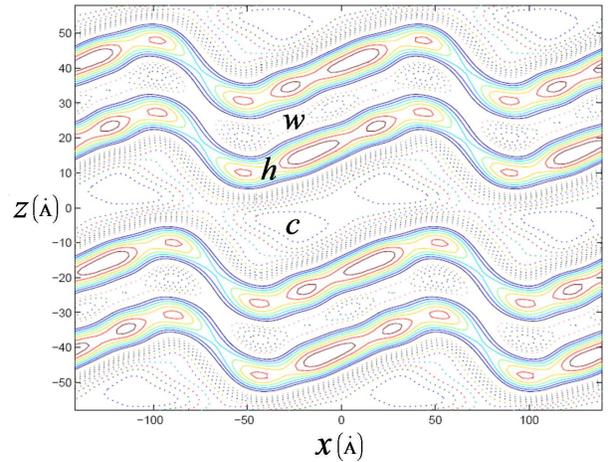}
\end{center}
\caption{Electron density map of the ripple phase of dimyristoylphosphatidylcholine calculated from x-ray diffraction
data \cite{kheya-thesis}. Bands labeled $h$ and $c$ correspond to the headgroup and
hydrocarbon chain regions of the bilayer; $w$ denotes the water layer separating the bilayers. }
\label{ripple_edm}
\end{figure}

Extensive studies using a variety of experimental 
techniques \cite{Tardieu73,Wack88,Hentschel91,Katsaras95,Sun96,Kheya03, 
Sackmann88, Luna77,Sackmann79,Hicks87,Zasadzinski88,Meyer96, Hansma88,Woodward97}, reveal that the $P_{\beta^{'}}$ phase 
is characterized by a periodic saw-tooth height modulation of the bilayers having an amplitude 
of $\sim$ 1 nm and a wavelength of $\sim$ 15 nm, and a  bilayer thickness that  is different 
in the two arms of the ripple (fig. \ref{ripple_edm}) \cite{Sun96,Kheya03}.
As a result, the rippled bilayers lack a mirror plane normal to the rippling direction.
While in  principle, this discrete symmetry breaking can arise from an asymmetry in either \emph{shape} (unequal
lengths of the two arms) or \emph{bilayer thickness} (unequal bilayer thickness in the two arms),
in practice these asymmetries seem to appear simultaneously. 

At first it was believed that the origins of the asymmetric ripple lay in the chirality of lipid molecules
\cite{Lubensky93}. However, subsequent experiments using racemic mixtures showed this was not the case
\cite{Zasadzinski88,Katsaras95}.  More recently, all-atom molecular dynamics simulations of lipid bilayers have observed
that the degree of chain ordering is different in the two arms of the ripple \cite{deVries05}. The occurrence of the
ripple phase only in those lipids that exhibit a $L_{\beta^{'}}$ phase at lower temperatures \cite{Watts78}, and in
isolated bilayers \cite{Mason99}, suggests an intimate connection between chain tilt and the ability of the bilayers to
form ripples.

Several theoretical models have been proposed to describe the sequence of phase transitions in such 
lipid bilayers and the microstructure of the ripple phase ~\cite{Doniach79,
Falkovitz82, Marder84, Sethna87, Leibler88, Leibler89, Kimura91, Lubensky93, Lubensky95, Seifert96, Kheya01,Komura08}.  
None of them accounts for all the observations. We list three 
key features that should be explained by any theory of the {\it ripple phase in achiral bilayers} 
: (1)  occurrence of $P_{\beta^{'}}$ phase between 
 $L_\alpha$ and $L_{\beta^\prime}$ phases, separated by two first-order transitions; (2) 
unequal bilayer  thickness in the two arms of the ripple; and (3) unequal lengths of the two arms.

\begin{figure}[h]
\begin{center}
\includegraphics[width= 85 mm,angle=0] {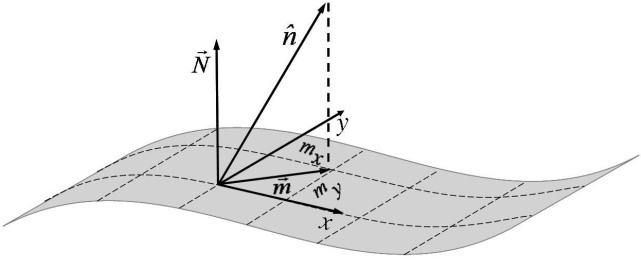}
\end{center}
\caption{The unit vector $\hat{\bf{n}}$ represents the orientation of the long axis of the lipid molecules relative to
the bilayer normal $\vec{\bf{N}}$. $\vec{\bf{m}}=
\vec{\bf{n}}-(\vec{\bf{N}}\cdot \hat{\bf{n}})\:\hat{\bf{N}}$ is the  projection of $\hat{\bf{n}}$ on the bilayer plane.}
\label{bilayer_m}
\end{figure}

In this paper we present a phenomenological Landau theory to describe the ripple phase in an isolated, achiral lipid
bilayer. Our free energy expression is written in
terms of two order parameters: a scalar order parameter $\psi$ and a  2-D vector order 
parameter ${\bf m}$ (fig.\ref{bilayer_m}).  
$\psi$ describes the melting of the bilayer and is the difference in the bilayer 
thickness ~\cite{Leibler88,Leibler89} between the fluid ($L_\alpha$) and
ordered ($P_{\beta^{'}}$ and $L_{\beta^{'}}$) phases. Since the bilayer thickness is determined by the conformations of 
the hydrocarbon chains,
$\psi$ can also be interpreted in terms of differences in chain conformations between the fluid and ordered phases 
\cite{Kimura91}. $\bf{m}$ is the projection of the molecular axis on the bilayer 
plane
~\cite{Lubensky93}. A third order parameter $h$, describing the height of the bilayer, can be integrated out of the 
expression for
the total free energy density. This model is found to capture all the three salient features of the ripple
phase listed above.

The total free energy per unit area is taken to be the sum of three terms; the stretching free energy density $f_s$, the
tilt free energy density $f_t$, and the curvature free energy density $f_c$. For an isolated lipid bilayer $f_s$ is
given by~\cite{Komura08},
\begin{align}
f_{s}=&\frac{1}{2}\:a_2\:\psi^2\:+\:\frac{1}{3}\:a_3\:\psi^3\:+\:\frac{1}{4}\:a_4\:\psi^4\:
+\:\frac{1}{2}C\left(\nabla\psi\right)^2\nonumber
\\ &+\:\frac{1}{2}D\left(\nabla^2\psi\right)^2+\:\frac{1}{4}E\:\left(\nabla\psi\right)^4 
\end{align}
where $\psi(x,y)\:=\:\frac{\delta(x,y)\:-\:\delta_0}{\delta_0}$, $\delta(x,y)$ being the membrane
thickness at position $(x,y)$ in the bilayer measured with respect to a flat reference plane, and
$\delta_0$ the constant thickness of the membrane in the
$L_\alpha$ phase. $\psi$ is taken to be positive for $T$ $<$ 
$T_m$ due to the 
stretching of the 
chains. This is valid in general, even if the chains are tilted below $T_m$.
Explicit temperature dependence is assumed to reside solely in the coefficient of $\:\psi^2$ :
$a_2\:=\:a_2^{\prime}\:(T-T^*)$, $T^*$ being a reference temperature.
$a_3$ is taken to be negative, so that the continuous transition at $T^*$ is
preempted by a first order melting transition at $T_m\;=\;T^*\;+\;\frac{2 a_3^{2}}{9a_2^{\prime} a_4}$. The coefficient 
C
can either be positive or negative, but $a_4$, $D$ and $E$  are always positive to ensure stability. 
With $C>0$, the equilibrium phases are always homogeneous in space; either as $L_\alpha$ or 
$L_{\beta}$ ($L_{\beta^{'}})$.
However, with $C<0$ modulated phases are possible with some characteristic wave vector $q_0$.
The $\left(\nabla\psi\right)^4$ term is included, since in the context of a one dimensional model with a scalar order
parameter, it has been shown that such a term is necessary to stabilize a 
modulated phase with a non zero mean value of the order parameter ~\cite{Jacobs84}.

The tilt free energy density can be written as, 
\begin{align}
f_{t}=& \frac{1}{2}b_{2}\left|\textbf{m}\right|^2+ \frac{1}{4}b_{4}\left|\textbf{m}\right|^4+
\tilde{\Gamma}_1\left(\nabla\cdot\textbf{m}\right)^2 \nonumber\\
&+\Gamma_2\left(\nabla^2\textbf{m}\right)^2 +\Gamma_3\left(\nabla\cdot\textbf{m}\right)^4
+\Gamma_{4}\psi\left|\textbf{m}\right|^2 \nonumber\\
&+\Gamma_{5}\left(\textbf{m}\cdot\nabla\psi\right)^2+\Gamma_{6}\left(\textbf{m}\times\nabla\psi\right)^2\nonumber\\
&+\Gamma_{7}\left(\nabla^{2}\psi\right)\left(\nabla\cdot\textbf{m}\right)^2
+\Gamma_{8}\left(\nabla\psi\right)^2\left(\nabla\cdot\textbf{m}\right)^2
\label{f_t}
\end{align} The first four terms in eqn.(\ref{f_t}) are the usual terms in the expansion of the free energy in terms of 
a vector 
order parameter. The 
$\left(\nabla\cdot\textbf{m}\right)^4$ term is
included to be consistent with the $\left(\nabla\psi\right)^4$ term introduced in $f_s$.
The next term represents the coupling between $\psi$ and ${\bf m}$, which is responsible for the appearance of
tilted phases in this model, as $b_2$ is taken to be positive. If $\Gamma_4$ $>$ 0, the stable phase below $T_m$ is 
$L_\beta$ with $\left|\textbf{m}\right|$=0. On the other hand, tilted phases can form if
$\Gamma_4$ $<$ 0. The succeeding two terms take into account the anisotropy of the tilted bilayer. 
The next two terms represent higher order couplings between modulations in
$\psi$ and in $\bf{m}$, allowed by the symmetry of the system.

\begin{figure}[h]
\begin{center}
\includegraphics[width=90mm,angle=0] {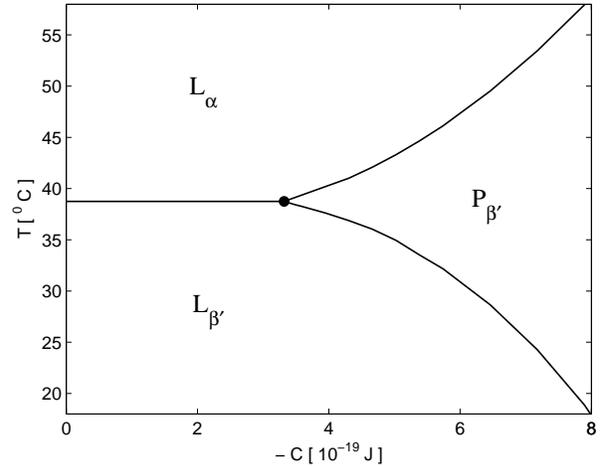}
\end{center}
\caption {Phase diagram in the $T-C$ plane calculated from the model. $a_2^\prime$ = 159.42 $k_B$, $T^*$ = 260.0 K.
Values of the  other coefficients in units of $k_BT^*$ are: $a_3$ = -306.5, $a_4$ = 613.15, $b_2$ = 0.2, $b_4$ = 200.0, 
$D$ = 557.41, $E$ = 600.0, $\Gamma_1$ = 0.010, $\Gamma_2$ = 1.80, $\Gamma_3$ = 500.0, 
$\Gamma_4$ = -3.0, $\Gamma_5$ = -20.0, $\Gamma_6$ = -20.0, $\Gamma_7$ = -500.0, $\Gamma_8$ = -750.0. Both the 
main-transition ($L_\alpha$ $\rightarrow$ $L_{\beta^\prime}$; $L_\alpha$ $\rightarrow$ $P_{\beta^\prime}$) and 
pre-transition ($P_{\beta^\prime}$ $\rightarrow$ $L_{\beta^\prime}$) are first order.} 
\label{phasedia}
\end{figure}

\begin{figure}[h]
\begin{center}
\includegraphics[width=90mm,angle=0] {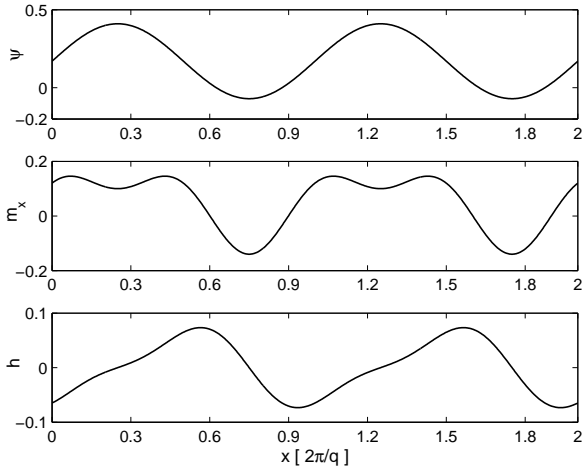}
\end{center}
\caption{Spatial variation of the different order parameters in the $P_{\beta^\prime}$ phase at $T$ = 310.5 K and $C$= 
-4.84 $\times 10^{-19}$ J. }
\label{ops}
\end{figure}

\begin{figure}[h]
\begin{center}
\includegraphics[width=80mm,angle=0] {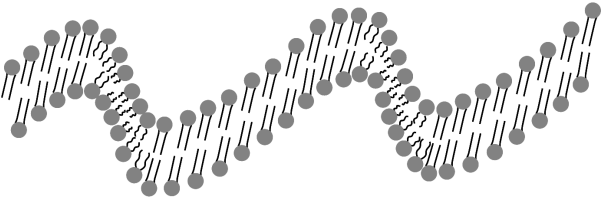}
\end{center}
\caption{Schematic of the bilayer profile obtained from the model. The bilayer thickness is different in the two arms 
of the saw-tooth-like ripples.} 
\label{schm}
\end{figure}

The curvature energy density of the bilayer can be written as  \cite{Lubensky93, Lubensky95},
\begin{equation}
f_c\:=\:\frac{1}{2}\:\kappa\:\left(\nabla^2\:h\right)^2\:-\:\gamma\left(\nabla^2\:h\right)\:\left(\nabla\cdot\textbf{m}\right)
\end{equation}
where $h(x,y)$ is the height of the bilayer relative to a flat reference plane, $\kappa$ is the bending rigidity of the
membrane, and  $\gamma$ couples the mean curvature to splay in \textbf{m}.

The equilibrium height profile of the bilayer $h(x,y)$ is related to the tilt \textbf{m} via the Euler-Lagrange 
equation,
\begin{equation}
\nabla^2\:h\:=\:\frac{\gamma}{\kappa}\:\left(\nabla\cdot\textbf{m}\right)
\end{equation}
Eliminating $h$ from the free energy density $f$ leads to the effective energy density $f_{eff}$ with a reduced splay
elastic constant $\Gamma_1=\tilde\Gamma_1 - \gamma^2/(2\kappa)$.

To determine the mean field phase diagram we choose the following ansatz for $\psi$ and
$\textbf{m}$,
\begin{align}
\psi = & ~\psi_0+\:\psi_1sin(qx) \nonumber \\
m_x = & ~m_{0x}+\:m_{1x}\:cos\left(qx\right)+m_{2x}\:sin\left(qx\right)\nonumber\\
& +\:m_{3x}\:cos\left(2qx\right)+m_{4x}\:sin\left(2qx\right) \nonumber\\
m_y  = & ~m_{0y}
\end{align}

We do not consider two-dimensionally modulated ripples, since they do not appear to be generic; there being, as far as 
we know, only one report of such a structure \cite{Yang05}. Spatial modulation of $m_y$ is 
neglected as we do not keep terms proportional to $(\nabla\times\bf{m})$ in eqn.(\ref{f_t}) for 
reasons discussed below. Higher order Fourier components of $m_x$ are retained in order to account for the ripple 
asymmetry. Three different ripple structures can be described with this ansatz: (1) the $P_\beta$ with no 
mean tilt ($m_{0x}$ = $m_{0y}$ = 0), (2) the $P_{\beta^\prime}^T$ with a mean tilt along $y$ ($m_{0x}$ = 0, $m_{0y}$ 
$\neq$ 0) , and 
(3) the $P_{\beta^\prime}^L$ with a mean tilt along $x$ ($m_{0x}$ $\neq$ 0, $m_{0y}$ = 0). Of these only the 
$P_{\beta^\prime}^L$ structure has an asymmetric height profile \cite{Kheya01}.

The phase diagram in the C-T plane obtained from numerical minimization of the effective free energy density 
averaged over one spatial period,  $<f_{eff}>$ = $(q/2\pi) \int_{0}^{2\pi/q} f_{eff}\:dx $, is given in 
fig. \ref{phasedia}. It is calculated for a choice of parameter values, which 
reproduce closely the main- and pre-transition temperatures of dipalmitoylphosphatidylcholine  at $C$ = - 
4.84$\times10^{-19}J$. The values 
of $T^*$, $a_2$, $a_3$,$a_4$, and $D$ are similar to those used in 
refs.~\cite{Leibler89} and \cite{Komura08}. As can be seen from fig. \ref{phasedia}, there are three 
distinct regions in
the phase diagram corresponding to three different phases:  $L_{\alpha}~(\psi_0 = \psi_1 = m_{0x} = m_{0y} = m_{1x} = 
m_{2x} =  m_{3x} = m_{4x} = 0)$, $L_{\beta^{'}}~(\psi_0\neq0,  m_{0x}\neq0, m_{0y}\neq0;
\psi_1 = m_{1x} = m_{2x} = m_{3x} = m_{4x} = 0)$, and $P_{\beta^\prime}^L~ (\psi_0\neq0,\psi_1\neq0, m_{0x}\neq0,  
m_{2x}\neq0, m_{3x}\neq0; m_{1x} = m_{4x} =  m_{0y} = 0)$.
\par
It is interesting that of the three possible ripple structures only the asymmetric $P_{\beta^\prime}^L$, which is 
similar to the experimentally observed structure, is present in 
this phase diagram \footnote{We have not exhaustively scanned the parameter space to see if the other two structures are 
stable in some other regions.}.
The first-order transition lines which separate these three phases meet at the Lifshitz point located at
$C_{Lp}= - 3.32 \times10^{-19}J$ and $T = 38.75^{\circ}C$. For $C > C_{Lp} $, the first order 
$L_{\beta^{'}}$ $\rightarrow$ $L_{\alpha}$ transition line  is parallel to
the C axis and occurs at $T_m=38.75^{\circ}C$.  But
for $C < C_{Lp}$, the intermediate $P_{\beta^{'}}$ phase is found, separated from the other 
two phases by first-order transition lines. Further,
the region occupied by the $P_{\beta^{'}}$ phase expands at the expense of the other two as $C$ becomes more negative. 

Typical spatial variation of the order parameters in the ripple phase is shown in fig. \ref{ops}. The height profile is
asymmetric and resembles very closely those seen in experiments (fig. \ref{ripple_edm}) \cite{Woodward97,Sun96,Kheya03}.  
$\psi_1$ is almost $\pi$/2 out of phase with $h$, so that it is positive (negative) along the longer (shorter) arm of 
the ripple, resulting in different bilayer thicknesses
in the two arms, again in agreement with experimental observations (fig. \ref{ripple_edm}) \cite{Sun96,Kheya03}.  Fig. 
\ref{schm}  shows a 
schematic
of the structure of the bilayer inferred from these results. It is clear that the model presented 
here accounts for all the salient features of the ripple phase listed in the introduction.

The essential term in the free energy expression responsible for the asymmetric ripples is the one with the coefficient
$\Gamma_8$, since similar structures can be obtained by setting $\Gamma_5$=$\Gamma_6$=$\Gamma_7$ = 0, as long as
$\Gamma_8$ is $<$ 0. Thus the present model spontaneously picks out a non-zero value of the mean tilt along the
rippling direction $q$, even in the absence of any explicit in-plane anisotropy of the bending rigidity. This is in
contrast to the model presented in refs. \cite{Lubensky93} and \cite{Lubensky95}, where the mean tilt occurs in a
direction normal to $q$, resulting in symmetric ripples in the case of achiral bilayers; the bending rigidity has to be
explicitly taken to be lower along the tilt direction in order to obtain a non-zero mean tilt along $q$ and to stabilize
asymmetric ripples within this model \cite{Kheya01}. If $\Gamma_7$ is made sufficiently positive in the present model,
$\psi_1$ and $h$ become almost in phase, so that the bilayer thickness is modulated within each arm of the ripple.
It might be possible to tune this parameter by a suitable choice of an impurity which would prefer to smoothen  
variations in $\psi$; in such cases we predict the existence of this new ripple phase. 

We have included only terms proportional to $(\nabla\cdot \bf{m})$ in the expression for $f_t$. In
general there will also
be terms proportional to $(\nabla \times \bf{m})$, which lead to a ripple structure with a  non-zero winding number in
the model presented in refs. \cite{Lubensky93} and \cite{Lubensky95}.
However, such a
structure can be expected to be energetically very unfavorable in an achiral bilayer, since it is not consistent with
parallel
close-packing of the chains demanded by van der Waals interaction.

A straightforward extension of this model would be to use a better description of the chain-melting transition, instead 
of the reduced bilayer thickness $\psi$ employed here. Possible order parameters include components of the
in-plane density wave, as considered in the theory of weak crystallization \cite{Kats93}, and those of herringbone 
order, used to describe positional ordering in monolayers \cite{Kaganer99}. However, such an attempt would be useful 
only if the details of chain ordering in this phase, presently unknown, can be experimentally established. 

%\acknowledgments

We thank Yashodhan Hatwalne for many valuable discussions and Kheya Sengupta for the electron density map of the ripple 
phase.


\begin{thebibliography}{99}

\bibitem{Tardieu73}
A.\ Tardieu, V.\ Luzzati and F.\ C.\ Reman, J. Mol. Biol. {\bf{75}}, 711 (1973);

\bibitem{Smith90} G.\ B.\ Smith, E.\ B.\ Sirota, C.\ R.\ Safinya, and N.\ A.\ Clark, Phys. Rev. Lett. {\bf 60}, 813
(1988);  G.\ B.\ Smith, E.\ B.\ Sirota, C.\ R.\ Safinya, R.\ J.\ Plano, and N.\ A.\ Clark, J. Chem. Phys. {\bf 92}, 4519
(1990)

\bibitem{Sun94} W.\ -J.\ Sun, R.\ M.\ Sutter, M.\ A.\ Knewtson, C.\ R.\ Worthington, S.\ Tristam-Nagle, R.\ Zhang, and
J.\ F.\ Nagle, Phys. Rev. E. {\bf 49}, 4665 (1994).

\bibitem{MacIntosh80}
T.\ J.\ MacIntosh, Biophys. J. {\bf 294}, 237 (1980).

\bibitem{Janiak79}
M.\ J.\ Janiak, D.\ M.\ Small, and G.\ G.\ Shipley, J. Mol. Biol. {\bf 254}, 6068 (1979).

\bibitem{Wack88}
D.\ C.\ Wack and W.\ W.\ Webb, Phys. Rev. Lett. {\bf 61}, 1210 (1988).
                                                                                                                              
\bibitem{Hentschel91}
M.\ P.\ Hentschel and F.\ Rustichelli, Phys. Rev. Lett. {\bf 66}, 903 (1991).

\bibitem{Katsaras95}
J.\ Katsaras and V.\ A.\ Raghunathan, Phys. Rev. Lett. {\bf{74}}, 2022 (1995).

\bibitem{Sun96} W.\ -J.\ Sun, S.\ Tristam-Nagle, R.\ M.\ Sutter, and
J.\ F.\ Nagle, Proc. Natl. Acad. Sci. {\bf 93}, 7008 (1996).

\bibitem{Kheya03}
K.\ Sengupta, V.\ A.\ Raghunathan, and J.\ Katsaras, Phys. Rev. E{\bf 68}, 031710 (2003).

\bibitem{Sackmann88}
K.\ Mortensen, W.\ Pfeiffer, E.\ Sackmann, and W.\ Knoll, Biochim. Biophys. Acta {\bf 945}, 221 (1988).

\bibitem{Luna77}
E.\ J.\ Luna and H.\ M.\ McConnell, Biochim. Biophys. Acta {\bf 470}, 303 (1977).


\bibitem{Sackmann79}
R.\ Krbecek, C.\ Gebhardt, H.\ Gruler, and E.\ Sackmann, Biochim. Biophys. Acta {\bf 554}, 1 (1979).

\bibitem{Hicks87}
A.\ Hicks, M.\ Dinda, and M.\ A.\ Singer, Biochim. Biophys. Acta {\bf 903}, 177 (1987).
                                                                                                                              
\bibitem{Zasadzinski88}
J.\ A.\ N.\ Zasadzinski, Biochim. Biophys. Acta {\bf 946}, 235 (1988).
                                                                                                                              
\bibitem{Meyer96}
H.\ W.\ Meyer, Biochim. Biophys. Acta {\bf 1302}, 138 (1996).
                                                                                                                              
\bibitem{Hansma88}
J.\ A.\ N.\ Zasadzinski, J.\ Scheir, J.\ Gurley, V.\ Elings, and P.\ K.\ Hansma, Science {\bf 239}, 1013 (1988).
                                                                                                                              
                                                                                                                              
\bibitem{Woodward97}
J.\ T.\ Woodward and J.\ A.\ Zasadzinski, Biophys. J. {\bf 72}, 964 (1997).

\bibitem{kheya-thesis} 
Kheya Sengupta, Ph.D. thesis, Jawaharlal Nehru University (2000).

\bibitem{Lubensky93}
T.\ C.\ Lubensky and  F.\ C.\ Mackintosh, Phys. Rev. Lett. {\bf71}, 1565 (1993).
                                                                                                                              
\bibitem{deVries05}
A. H.\ de Vries, S. Yefimov, A. E. Mark, and S. J. Marrink, Proc. Natl. Acad. Sci. {\bf 102}, 5392 (2005).

\bibitem{Watts78} A. Watts, K. Harlos, W. Maschke, and D. Marsh, Biochim. Biophys. Acta {\bf 510}, 63   (1978)

\bibitem{Mason99}
P.\ C.\ Mason, B. D. Gaulin, R. M. Epand, G. D. Wignall, and J. S. Lin, Phys. Rev. E {\bf 59}, 3361 (1999).


\bibitem{Doniach79}
S.\ Doniach, J. Chem. Phys. {\bf{70}}, 4587 (1979).

\bibitem{Falkovitz82}
M.\ S.\ Falkovitz, M.\ Seul, H.\ L.\ Frisch, and H.\ M.\ McConnell, Proc. Natl. Acad. Sci. {\bf 79}, 3918 (1982).

\bibitem{Marder84}
M.\ Marder, H.\ L.\ Frisch, J.\ S.\ Langer, and H.\ M.\ McConnell, Proc. Natl. Acad. Sci. {\bf 81}, 6559 (1984).

\bibitem{Sethna87}
J.\ M.\ Carlson and  J.\ P.\ Sethna, Phys. Rev. A {\bf 36}, 3359 (1987).

\bibitem{Leibler88}
R.\ E.\ Goldstein and S.\ Leibler, Phys. Rev. Lett. {\bf{61}}, 2213 (1988).

\bibitem{Leibler89}
R.\ E.\ Goldstein and S.\ Leibler, Phys. Rev. A. {\bf{40}}, 1025 (1989).

\bibitem{Kimura91}
K.\ Honda and H.\ Kimura, J. Phys. Soc. Jpn. {\bf 60}, 1212 (1991).

\bibitem{Lubensky95}
C.\ -M.\ Chen, T.\ C.\ Lubensky, and  F.\ C.\ Mackintosh, Phys. Rev. E {\bf51}, 504 (1995).

\bibitem{Seifert96}
U.\ Seifert, J\. Shillcock, and P.\ Nelson, Phys. Rev. Lett. {\bf 77}, 5237 (1996).

\bibitem{Kheya01}
K.\ Sengupta, V.\ A.\ Raghunathan, and Y.\ Hatwalne, Phys. Rev. Lett. {\bf 87}, 055705 (2001).

\bibitem{Komura08}
N.\ Shimokawa, S.\ Komura, and D.\ Andelman, Eur. Phys. J. E {\bf 26}, 197 (2008).


\bibitem{Jacobs84}
A.\ E.\ Jacobs, C.\ Grein, and F.\ Marsiglio, Phys. Rev. B {\bf 29}, 4179 (1984).

\bibitem{Kheya99}
K.\ Sengupta, V.\ A.\ Raghunathan, and J.\ Katsaras, Phys. Rev. E {\bf 59}, 2455 (1999).


\bibitem{Yang05}
L.\ Yang and M.\ Fukuto, Phys. Rev. E{\bf 72}, 010901 (2005).


\bibitem{Kats93} E. I. Kats, V. V. Lebedev, and A. R. Muratov, Phys. Rep. {\bf 228}, 1 (1993).

\bibitem{Kaganer99} V.\ M.\ Kaganer, H.\ M\:{o}wald, and P.\ Dutta, Rev. Mod. Phys. {\bf 71}, 779 (1999).

\end{thebibliography}
\end{document}